\documentclass[twocolumn]{aastex631}

\begin{document}

\title{Where are the Population III star relics in the simulated Milky Way analogues?}

\author[0000-0003-3279-0134]{Hang Yang}
\affiliation{School of Physics and Laboratory of Zhongyuan Light, Zhengzhou university}
\affiliation{National Astronomical Observatories, Chinese Academy of Sciences, Beijing 100101, China}
\affiliation{School of Astronomy and Space Science, University of Chinese Academy of Sciences, Beijing 100049, China}
\email{E-mail: hyang@nao.cas.cn, lgao@bao.ac.cn}

\author[0009-0006-3885-9728]{Liang Gao}
\affiliation{School of Physics and Laboratory of Zhongyuan Light, Zhengzhou university}
\affiliation{Institute for Frontiers in Astronomy and Astrophysics, Beijing Normal University, Beijing 102206, China}
\affiliation{National Astronomical Observatories, Chinese Academy of Sciences, Beijing 100101, China}

\author[0000-0002-7972-3310]{Qi Guo}
\affiliation{National Astronomical Observatories, Chinese Academy of Sciences, Beijing 100101, China}
\affiliation{School of Astronomy and Space Science, University of Chinese Academy of Sciences, Beijing 100049, China}
\affiliation{Institute for Frontiers in Astronomy and Astrophysics, Beijing Normal University, Beijing 102206, China}

\author[0000-0002-0389-9264]{Haining Li}
\affiliation{National Astronomical Observatories, Chinese Academy of Sciences, Beijing 100101, China}

\author[0000-0001-8382-6323]{Shi Shao}
\affiliation{National Astronomical Observatories, Chinese Academy of Sciences, Beijing 100101, China}

\author[0000-0002-8980-945X]{Gang Zhao}
\affiliation{National Astronomical Observatories, Chinese Academy of Sciences, Beijing 100101, China}
\affiliation{School of Astronomy and Space Science, University of Chinese Academy of Sciences, Beijing 100049, China}

\begin{abstract}

Using 6 Milky Way analogues with two different numerical resolutions from the Auriga simulation, we investigate the total mass, spatial distribution and kinematics of the Population III star relics in the Milky Way analogues at $z=0$. These relics (primarily second generation stars) formed over a wide redshift range, from about $z=22$ to $z=4$, with an average formation redshift of $z \sim 10.0$, and comprise about $2\times10^{-5}$ of the entire galactic stellar population. The disk and bulge components host only a small fraction of these relics, contributing less than $12$ percent in total. The stellar halo, in particular the outer stellar halo of which galactic radius $r>30$ kpc, hosts the largest fraction (about 46 percent on average), with an average of one relic star for per $4,000$ to $10,000$ stars, making it a promising region for observational searches. Additionally, around $18$ percent of the Population III star relics are found in satellite galaxies, with smaller and older satellite galaxies tending to contain a higher proportion of these stars. Thus, low-mass and early-formed satellite galaxies are also ideal targets for finding such relics, although some satellite galaxies may lack them entirely. The spatial distribution and kinematics of these stars show good numerical convergence across different simulation resolutions. Our results provide valuable guidance for searches of the Population III star relics and offer insights for interpreting findings from ongoing and future stellar archaeology surveys.
\end{abstract}

\keywords{Milky Way Galaxy (1054); Galactic archaeology (2178); Population II stars (1284); Hydrodynamical simulations (767)}

\section{Introduction} \label{sec:intro}
The first generation stars (also often referred to as population III stars, or Pop III stars) are thought to have been born from primordial gas \citep[e.g.][]{2002Sci...295...93A, 2003ApJ...592..645Y, 2006ApJ...652....6Y, 2005MNRAS.363..393R, 2007MNRAS.378..449G} in the early Universe about a few hundred Myr after the Big Bang. In the standard $\Lambda$CDM paradigm, primordial gas could condense in dark matter halos with a virial mass about of $10^6\ \rm{M_{\odot}}$ via molecular cooling \citep[e.g.][]{1997ApJ...474....1T}, and then form the first stars once their density became sufficiently high. The formation of Pop III stars ends the dark ages of the Universe, prompts chemical enrichment, and initiates the cosmic reionization process. Although such classical formation scenarios have been proposed \citep[][for an overview]{2004ARA&A..42...79B, 2011ARA&A..49..373B, 2023ARA&A..61...65K}{}, properties of the Pop III stars remain uncertain. For example, their initial mass function (IMF) is still highly uncertain.

Although the statistical properties of the first generation stars are shaped by many complex and competing physical processes (such as fragmentation, mergers, radiation feedback, etc.), it is generally believed that the Pop III stars have a top-heavy IMF \citep[e.g.][]{2016MNRAS.462.1307S, 2017MNRAS.470..898H}. In this scenario, most Pop III stars evolve rapidly and soon end their lives either by exploding as supernovae (SNe) or by collapsing directly into black holes, depending on their masses  \citep[e.g.][]{2002A&A...382...28S}. The SNe of Pop III stars would eject metals into the surrounding environment, enriching the gas and enabling it to re-collapse, thereby forming the second-generation stars. These bona fide second-generation stars are usually expected to be high carbon-enhanced, and possibly carry the unique elemental abundance patterns (e.g., the strong odd–even effect for pair-instability supernovae (PISNe) descendants) that could trace the properties of their progenitor Pop III stars \citep[e.g.][]{2002ApJ...567..532H, 2014Sci...345..912A, 2015MNRAS.454..659J}. One of the most feasible methods to constrain the properties of first generation stars is through stellar archaeology surveys within the Milky Way \citep[e.g.][]{2015MNRAS.447.3892H, 2018ApJ...857...46I, 2022ApJ...931..147L}. By comparing the theoretical nucleosynthetic yields of Pop III SNe with the chemical composition of surviving second-generation stars in the nearby Universe, it becomes possible to infer the properties of Pop III stars. For instance, \cite{2023Natur.618..712X} identified the chemical signature of PISNe in a metal-poor star in the Milky Way, indicating a supermassive Pop III star progenitor with an initial mass of approximately 260 $\rm{M_{\odot}}$. Such observations provide direct constraints on the properties of Pop III stars. In addition, some theories predict the long-lived low mass ($M_{*} < 0.8 M_{\odot}$) Pop III stars \cite[e.g.][]{2011ApJ...737...75G, 2014ApJ...792...32S, 2016ApJ...826....9I, 2018MNRAS.473.5308M}. The discovery of such relics will directly constrain the lower mass limit of the Pop III IMF. Since these Pop III star relics in the Milky Way are expected to be very rare, the total mass, spatial distribution and kinematics of these relics are critical information for stellar archaeology studies.

In the past decade, studies based on both semi-analytic methods \citep[e.g.][]{2006ApJ...653..285S, 2010MNRAS.403.1283G} and cosmological hydrodynamical simulations \citep[e.g.][]{2007ApJ...661...10B, 2017MNRAS.465.2212S, 2018MNRAS.480..652E, 
2021MNRAS.500.3750S,
2023MNRAS.519..483C, 2023MNRAS.525L.105S} have aimed to predict the distribution of the Pop III star relics in the Milky Way. Most of the cosmological hydrodynamical simulations, however, do not explicitly model the Pop III stars formation. Hence, the bona fide second-generation stars cannot be included in these simulations. Instead, they often adopt extremely metal poor stars (EMP stars; [Fe/H] $<-3$) as the proxies for the Pop III star relics. In these simulations, the enrichment of EMP stars originates from the stars formed in the atomic cooling halos (first galaxy halos) rather than the genuine Pop III stars, which were born in molecular cooling halos. In reality, accounting for enrichment from these stars would wipe out the Pop III abundance signatures in EMP stars and weaken the constraints on properties of Pop III stars \citep[e.g.][]{2015MNRAS.454..659J}. Moreover, some studies have shown that the metallicity of the Pop III star relics is not necessarily lower than [Fe/H] = -3 \citep[e.g.][]{2017MNRAS.465..926D, 2022ApJ...929..119M, 2023Natur.618..712X}. 

In this short paper, we make use of the Auriga simulation and adopt a more reasonable tracer for the Pop III star relics, to explore their spatial distribution and kinematics at present day. Our results might provide valuable guidance for hunting the Pop III star relics and interpreting results from ongoing and future surveys. The structure of this paper is as follows: in Section 2, we describe the numerical simulation and methodology. The main results are presented in Section 3, and we conclude with a summary of our findings in Section 4.

\section{Method}
\subsection{Cosmological simulation}

The Auriga project \citep[][]{2017MNRAS.467..179G} comprises a suite of zoom-in cosmological simulations of Milky Way-mass dark matter halos and their surroundings. The parent Auriga halos were selected from a dark matter only simulation of the EAGLE \citep[][]{2015MNRAS.446..521S} project. The Auriga projects were performed with the magneto-hydrodynamic moving mesh code \texttt{AREPO} \citep[][]{2010MNRAS.401..791S}, and the adopted cosmological parameters are $\Omega_m=0.307$, $\Omega_b=0.048$, $\Omega_{\Lambda}=0.693$, $h=0.6777$, $n_s=0.9611$ and $\sigma_8=0.829$ \citep[][]{2014A&A...571A..16P}. The basic information of the Milky Way analogues are shown in Table \ref{tab:sample}. The main implements of galaxy formation in Auriga include primordial and metal line cooling; a spatial uniform UV background that completed HI reionization at $z = 6$; a two phase ISM model; a stochastic star formation model in cell with gas number density larger than $n = 0.13\rm\ cm^{-3}$; stellar and AGN feedback. A more detailed description of the Auriga model can be found in \cite{2017MNRAS.467..179G}. 

Two different resolution runs, level-3 and level-4, are used in this paper, which correspond to the dark matter (baryonic) mass resolution of about $5\times{10^4}M_{\odot}\ (6\times{10^3} M_{\odot})$ and $4\times{10^5} M_{\odot}\ (5\times{10^4} M_{\odot})$, respectively. For brevity, we use L3 and L4 to refer to these two resolution runs in the following sections. The Friends-of-Friends \citep[][]{1985ApJ...292..371D} and SUBFIND \citep[][]{2001MNRAS.328..726S} algorithms were applied to construct halo and subhalo catalog. 

\subsection{The Pop III star relics sample}

\begin{table}
	\centering
	\caption{The general properties of the simulated central galaxy at $z=0$. From left to right, the columns are: 
(1) simulation name; 
(2) virial radius for host halo; 
(3) virial mass for host halo; 
(4) total stellar mass for central galaxy.
}
 
	\label{tab:sample}
	\begin{tabular}{lccccccc} 
		\hline
		Name & $R_{200}$ [kpc] & $\log_{10} M_{200}$ [${\rm M}_{\sun}$]  & $\log_{10} M_{*}$ [${\rm M}_{\sun}]$ \\
		\hline
		Au6-L4 & 213.8 & 12.02 & 10.72 \\
		Au6-L3 & 211.8 & 12.0 & 10.81 \\
  
        Au16-L4 & 241.5 & 12.18 & 10.83 \\
	    Au16-L3 & 241.5 & 12.18 & 10.96 \\
     	
        Au21-L4 & 238.6 & 12.16 & 10.91 \\
	    Au21-L3 & 236.7 & 12.15 & 10.94 \\

        Au23-L4 & 245.3 & 12.20 & 10.98 \\
        Au23-L3 & 241.5 & 12.18 & 10.95 \\

        Au24-L4 & 240.9 & 12.17 & 10.87 \\
        Au24-L3 & 239.6 & 12.17 & 10.94 \\

        Au27-L4 & 253.8 & 12.24 & 11.0 \\
        Au27-L3 & 251.4 & 12.23 & 11.0 \\

		\hline  
	\end{tabular}
\end{table} 

Due to the limitations of the numerical resolution and sub-grid physics models, most cosmological simulations are unable to resolve the formation of Pop III stars, including the Auriga project used in this work. Previous numerical studies have shown that the central regions of first galaxies have already been enriched by earlier Pop III stars \citep[e.g.][]{2010ApJ...716..510G, 2012ApJ...745...50W}, hence the very first stars formed in the first galaxies contain most direct information on properties of Pop III stars. While the gas physics in the first galaxy formation is not realistic in the Auriga simulation suites because the gas of such systems is in the un-enriched primordial state, the closest tracer of the Pop III star relics may be the metal free stars with [Fe/H] = 0 in the simulations as they are not polluted by any other star formation process in the first galaxies. Therefore, we select the metal free star particles as the Pop III star relics tracers in the Auriga simulation suites.

It is worth noting that our sample may include truly metal-free Pop III stars. The Lyman–Werner radiation could suppress Pop III star formation \citep[e.g.][]{2013MNRAS.428.1857J, 2016ApJ...823..140X} in the minihalo progenitors of an atomic cooling halo. Moreover, inefficient metal mixing processes \citep[e.g.,][]{2013ApJ...775..111P, 2017ApJ...834...23S} could allow metal free regions to persist within halos that have already been partially enriched by earlier Pop III stars. According to the above discussion, the selected metal-free star particles in the simulated first galaxies could plausibly correspond to truly Pop III stars. Although we are unable to determine their precise fraction in Auriga simulation, previous theoretical studies suggest that new metal-free Pop III star formation in atomic cooling halos is expected to be limited and subordinate \citep[e.g.][]{2016ApJ...823..140X, 2025ApJ...980...41B}.

For convenience and to avoid ambiguity, we refer to the Pop III star relics, which contain second-generation stars primarily and, if any, a limited number of truly Pop III stars, as PSR in the following sections.

\subsection{Orbital decomposition}

The kinematic features of such PSR samples provide important information for identifying them in observations. We adopt the total angular momentum direction of all stars within 10 kpc as the z-axis and then transform the position and velocity vectors of all star particles into the new galactic center coordinate system. Based on the galactic radius $r$ and circularity, defined as $\epsilon = j_{z}/rV_{c}(r)$, the orbit types of such PSR can be separated into four components, 1) the disc stars with $r<30$ kpc and $\epsilon>0.7$; 2) the bulge stars with $r<3$ kpc and $\epsilon<0.7$; 3) halo stars which include the inner halo stars with $r<30$ kpc and $\epsilon<0.7$; and 4) the outer halo stars with $r>30$ kpc. The similar decomposition methods have been adopted in previous works \citep[e.g.][]{2022A&A...660A..20Z, 2023MNRAS.519..483C}.

\section{Results}

\subsection{The formation time and total mass of PSR in the Milky Way analogues}

\begin{figure}
	\includegraphics[width=0.5\textwidth]{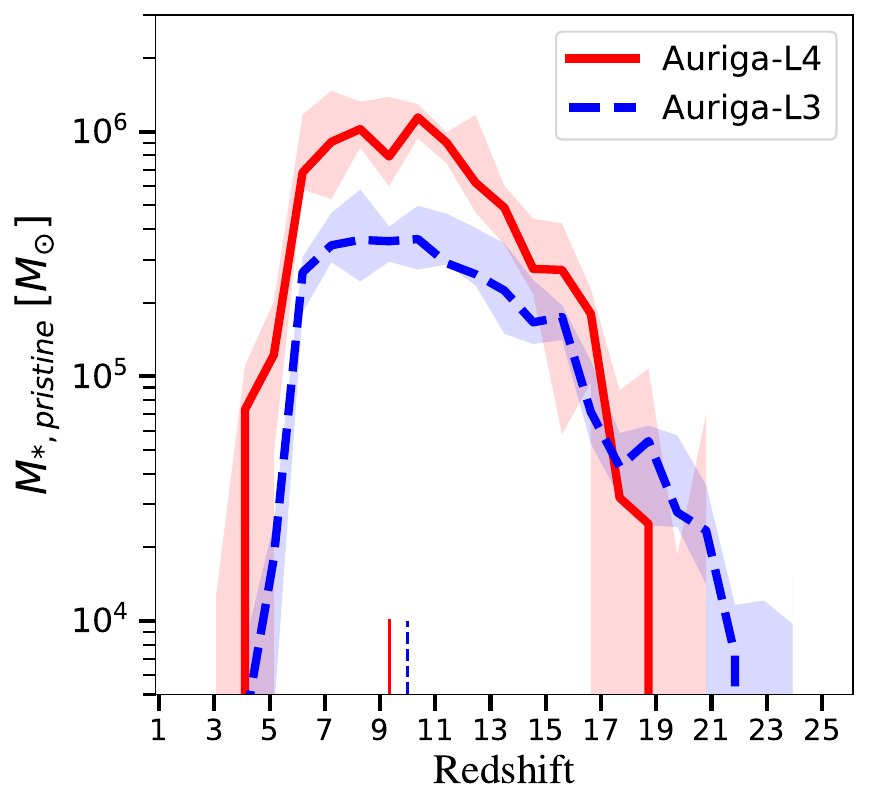}
	\centering
    \caption{The total mass of formed PSR as a function of formation redshift for six Milky Way analogues in the Auriga simulation suites. Results are presented for two different numerical resolutions. The red (blue) lines represent the median values of all six samples, while the shaded regions indicate the 16th–84th percentiles. The vertical lines represent the half-mass formation redshift of the PSR.}
    \label{fig:ft}
\end{figure}

It is interesting to see when and how many PSR have been formed in the Auriga simulations. In Fig.~\ref{fig:ft}, we plot the total mass of such stars which end up in the virial radius of 6 Auriga halos at $z=0$ as a function of redshift. The median values are shown and the shaded regions indicate the 16th-84th percentiles, different colors represent different resolution runs. Clearly, the formation epochs of the PSR span a wide redshift range, from $z\approx 22$ to $z\approx 4$. The average half-mass formation redshift for all PSR is about $z=10.0$ for the Auriga-L3 and $z=9.3$ for the Auriga-L4, as indicated by the vertical lines in the Fig~\ref{fig:ft}. Moreover, more than 85 (80) percent PSR have formed before $z=6$ for Auriga-L3 (L4). Intriguingly, some PSR could form as low as redshift about $z\approx 4$ in the simulations.

In total, the median mass of the PSR in six Auriga simulations at the present day is $10^{6.2}$ $(10^{6.6})\rm\ M_{\odot}$ for the Auriga-L3 (L4), accounting for about $2\ (5)\times10^{-5}$ of the entire galactic stellar population, making them extremely difficult to detect. Note that while the overall trends are similar between the L3 and L4 runs, the divergence in the quantitative results is expected between two different resolution runs because of numerical resolution effects and the adopted star formation model.

\subsection{The spatial distribution and kinematic of PSR in the Milky Way analogues}

\begin{figure*}
	\includegraphics[width=\textwidth]{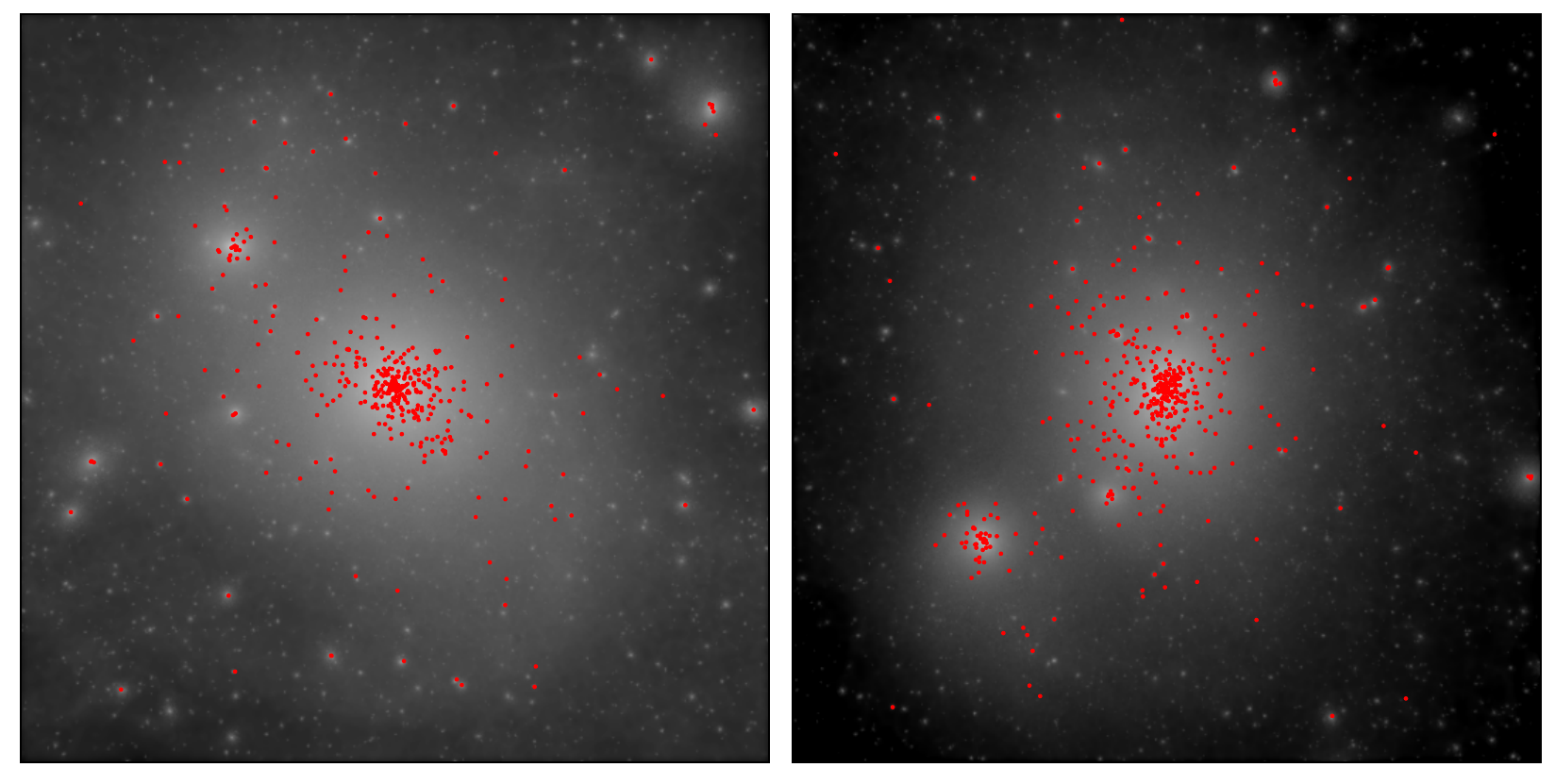}
	\centering
    \caption{The projected distribution of PSR (red dots) over underlying dark matter density fields of the Au6 (left) and Au16 (right) halos. The results are shown for the level 3 resolution runs. Each panel spans a physical scale of 300 kpc per side.}
    \label{fig:ill}
\end{figure*}

In order to have a visual impression of present day spatial distribution of the PSR, we overlay their positions on the underlying dark matter fields in the Au6-L3 and Au16-L3 runs, as shown in Fig.~\ref{fig:ill}. The PSR appear concentrated in the central regions of the halos, while a significant fraction is sparsely distributed throughout the stellar halo, and some reside in satellite galaxies. This distribution is a natural consequence of hierarchical structure assembly in the cold dark matter theory and is also consistent with the previous studies. The early-formed, low-mass first galaxies gave birth to these PSR. A significant fraction of these first galaxies were disrupted, contributing to the bulge and stellar halo, while some survived as present-day satellite galaxies, and still host their PSR.

\begin{figure}
	\includegraphics[width=\linewidth]{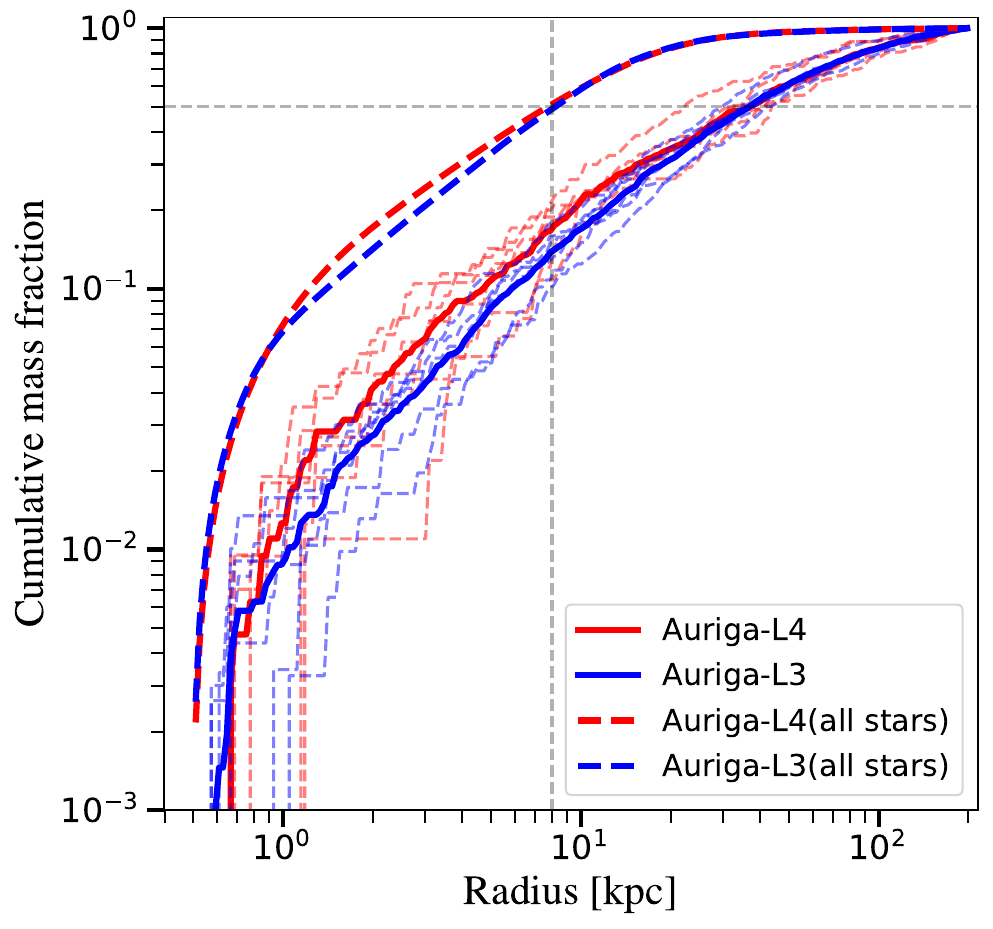}
	\centering
    \caption{The radial distribution of the PSR and all stars in Milky Way analogues at z=0. The bold solid lines represent the average radial distribution of the PSR, while the dashed lines indicate the average distribution of all stars, as specified in the legend. The light-colored lines correspond to results from individual simulated samples.}
    \label{fig:cdf}
\end{figure}

In Fig.~\ref{fig:cdf}, we show quantitative results for the radial spatial distribution of the PSR in the simulated Milky Way analogues. The bold solid blue and red lines represent the median radial cumulative distributions for Auriga-L3 and L4, respectively. For comparison, we also plot the median radial distribution of all stars in the Auriga simulation suites in bold dashed lines. Clearly, the PSR are substantially less concentrated than the overall stellar population: about half of these stars are located more than 40 kpc away from the galactic center, in stark contrast to the total galactic stellar population, where half of all stars are distributed within 10 kpc of the galactic center. While small differences between different resolution runs are evident in relatively inner regions, the overall numerical convergence is good.

\begin{figure*}
	\includegraphics[width=\textwidth]
    {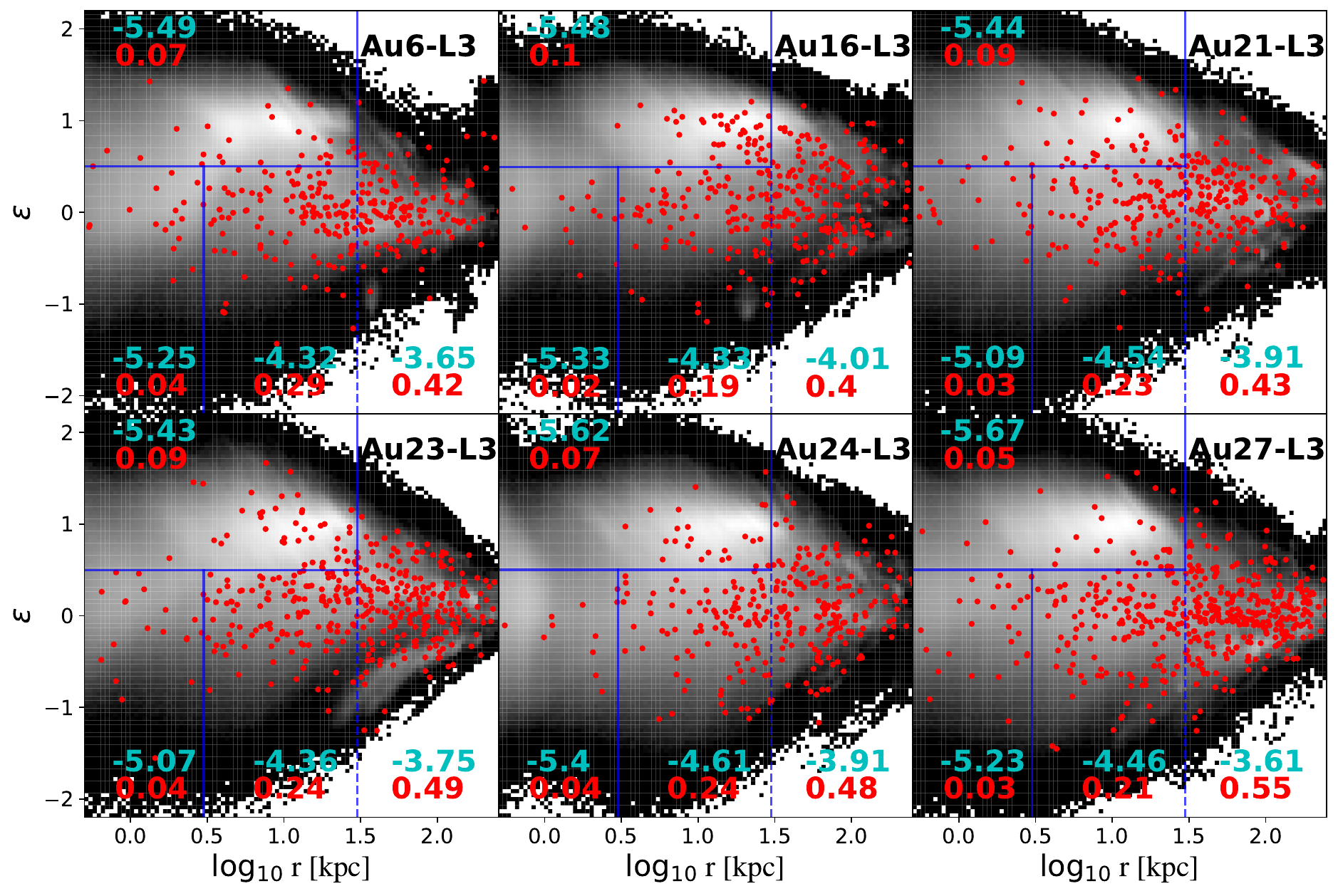}
	\centering
    \caption{The grey 2D histograms depict the orbital distribution of all stars in the central galaxy of the Auriga-L3, while the red dots represent the orbital properties of PSR. The regions outlined by blue lines correspond to different orbital types. The cyan numbers in each region indicate the mass ratio (logarithmic to base 10) of PSR with the corresponding orbital type to all stars, while the red numbers show the mass ratio of PSR with the corresponding orbital type to the total PSR population.}
    \label{fig:k3}
\end{figure*}

\begin{figure*}[htbp]
	\includegraphics[width=\textwidth]{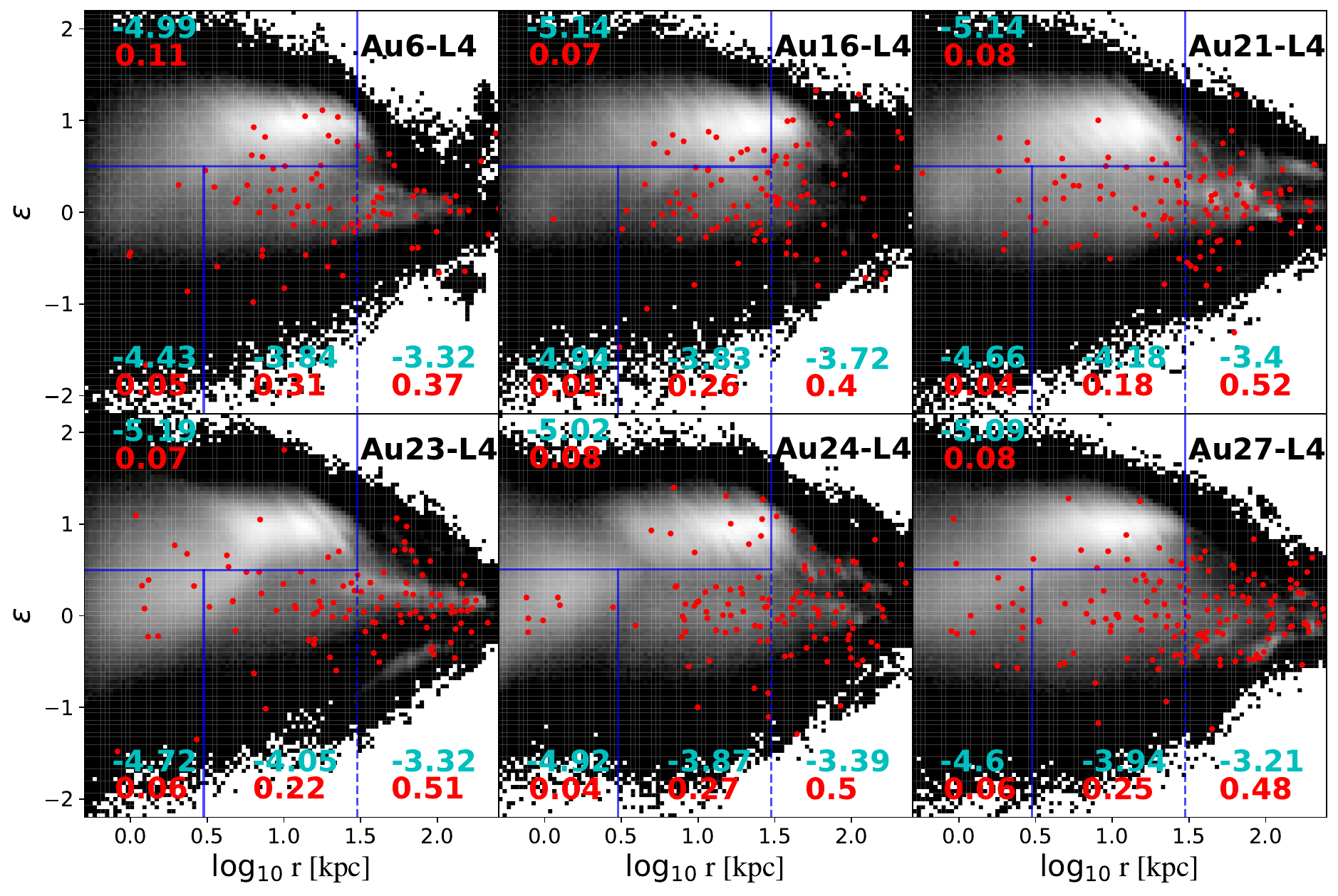}
	\centering
    \caption{Same as the content of Fig~\ref{fig:k3}, but it is the results for Auriga-L4.}
    \label{fig:k4}
\end{figure*}

We present the kinematically decomposed results for all stars and the PSR in six Milky Way analogues from the L3 and L4 simulations in Fig.\ref{fig:k3} and Fig.\ref{fig:k4}, respectively. Note that we exclude stars associated with satellite galaxies, which account for a significant fraction of the PSR. As shown in Table~\ref{tab:fraction}, this fraction varies from halo to halo, ranging from about 10 percent to 30 percent. The stellar orbit distribution $p(r, \epsilon)$ of all stars is shown for comparison, while the red dots represent the orbital properties of PSR in Fig.~\ref{fig:k3} and Fig.~\ref{fig:k4}. The red and cyan numbers in each panel show the ratio of PSR of the corresponding orbit type to all PSR and to all stars with the same orbit type, respectively. Note that the cyan numbers are expressed in logarithmic base $10$. For reference, these numbers are also listed in Table~\ref{tab:fraction}.

\begin{table*}
	\centering
	\caption{The kinematic decomposition of the PSR in each simulated central galaxy at $z=0$. From left to right, the columns are: 
1) PSR in bulge to total relic mass ratio $F_{B}$; 
2) PSR in disk to total PSR mass ratio $F_{D}$; 
3) PSR in inner halo to total PSR mass ratio $F_{iH}$; 
4) PSR in outer halo to total PSR mass ratio $F_{oH}$; 
5) PSR in satellite galaxies to total PSR mass ratio $F_{S}$; 
6) PSR in bulge to all bulge stars mass ratio $f_{B}$; 
7) PSR in disk to all disk stars mass ratio $f_{D}$; 
8) PSR in inner halo to all inner halo stars mass ratio $f_{iH}$; 
9) PSR in outer halo to all outer halo stars mass ratio $f_{oH}$.
}
 
	\label{tab:fraction}
	\begin{tabular}{lccccccccc} 
		\hline
		Name & $F_{B}$ & $F_{D}$ & $F_{iH}$ & $F_{oH}$ & $F_{S}$ & $\log_{10}f_{B}$ & $\log_{10}f_{D}$ & $\log_{10}f_{iH}$ & $\log_{10}f_{oH}$ \\
		\hline
        
		Au6-L3 & 0.04 & 0.07 & 0.29 & 0.42 & 0.18 & -5.25 & -5.49 & -4.32 & -3.65 \\
		Au6-L4 & 0.05 & 0.11 & 0.31 & 0.37 & 0.16 & -4.43 & -4.99 & -3.84 & -3.32 \\
  
        Au16-L3 & 0.02 & 0.10 & 0.19 & 0.40 & 0.29 & -5.33 & -5.48 & -4.33 & -4.01  \\
	    Au16-L4 & 0.01 & 0.07 & 0.26 & 0.40 & 0.23 & -4.94 & -5.14 & -3.83 & -3.72 \\
     	
        Au21-L3 & 0.03 & 0.09 & 0.23 & 0.43 & 0.22 & -5.09 & -5.44 & -4.54 & -3.91 \\
	    Au21-L4 & 0.04 & 0.08 & 0.18 & 0.52 & 0.18 & -4.66 & -5.14 & -4.18 & -3.40 \\

        Au23-L3 & 0.04 & 0.09 & 0.24 & 0.49 & 0.14 & -5.0 & -5.4 & -4.29 & -3.75 \\
        Au23-L4 & 0.06 & 0.07 & 0.22 & 0.51 & 0.14 & -4.72 & -5.19 & -4.05 & -3.32 \\

        Au24-L3 & 0.04 & 0.07 & 0.24 & 0.48 & 0.17 & -5.4 & -5.62 & -4.61 & -3.91 \\
        Au24-L4 & 0.04 & 0.08 & 0.27 & 0.5 & 0.11 & -4.92 & -5.02 & -3.87 & -3.39 \\

        Au27-L3 & 0.03 & 0.05 & 0.21 & 0.55 & 0.16 & -5.23 & -5.67 & -4.46 & -3.61 \\
        Au27-L4 & 0.06 & 0.08 & 0.25 & 0.48 & 0.13 & -4.60 & -5.08 & -3.94 & -3.21 \\
        
		\hline  
	\end{tabular}
\end{table*}

As shown in Fig.\ref{fig:k3} and Fig.\ref{fig:k4}, the stellar halo hosts the largest fraction of PSR, particularly the outer stellar halo, which contains about 46 (49) percent of them in Auriga-L3 (L4), respectively. The fraction of PSR in the bulge is the lowest, at only about 4 (5) percent in Auriga-L3 (L4). Taking into account the strong dust extinction and also high stellar density in the bulge, it is quite challenging to search for these stars observationally in this region. Consistent with expectation that ex-situ stars are generally believed to have pressure-supported orbits, the proportion of PSR in the disk is also quite low, at about 8 percent. These results suggest that the stellar halo is the most promising region to observe such stars in our Galaxy, especially in the outer stellar halo. According to the cyan numbers shown in Fig.~\ref{fig:k3}, there should be one relic star out of every 4,000 to 10,000 stars in the outer halo component of Milky Way analogues in Auriga-L3. In comparison, there should be only one relic star out of every approximately 100,000 disk stars.

As shown by the red numbers, the orbital distributions are roughly numerically converged in both resolutions. However, compared to the lower resolution runs, the ratio of PSR to all stars is slightly lower in the higher resolution runs, as indicated by the cyan numbers. This difference is due to the higher resolution simulations resulting in both a slightly larger total stellar mass and a smaller total PSR mass.

\subsection{The PSR in the satellite galaxies}

\begin{figure}[h]
	\includegraphics[width=0.45\textwidth]{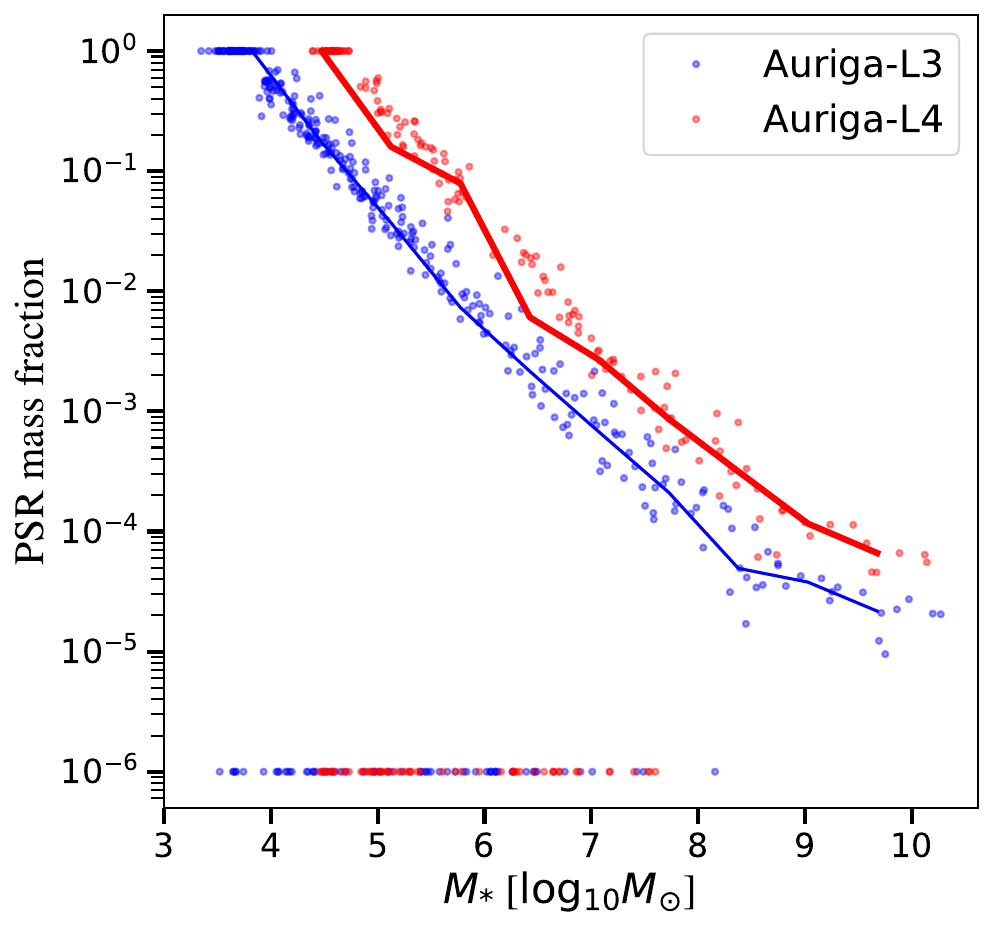}
	\centering
    \caption{The mass fraction of PSR relative to the total stellar mass ($M_{*}$) as a function of stellar mass for each surviving satellite galaxies in all six simulated Milky Way analogues at $z=0$. For satellite galaxies without PSR, their PSR mass fraction is arbitrarily set to $10^{-6}$.
}
    \label{fig:satellite}
\end{figure}

Previous studies suggest that the surviving satellite galaxies are also promising sites to hunt for the PSR \citep[e.g.][]{2015MNRAS.454..659J, 2017ApJ...848...85J}. As shown in the Table~\ref{tab:fraction}, about $18$ percent of the PSR mass budget is in satellite galaxies in Auriga-L3. To identify the best sites to search for the PSR, in Fig.~\ref{fig:satellite}, we plot the stellar mass $M_{*}$ of each simulated satellite galaxy against the mass ratio between the PSR and the total stars in the satellite. Results for both resolution runs are shown, with the solid line showing the median values. There is a strong correlation between $M_{*}$ and the PSR mass fraction, with less massive satellite galaxies containing a higher fraction of the PSR. In satellite galaxies with $M_{*}\sim 10^5 M_{\odot}$, the median mass fraction of the PSR is about $5$ percent, in stark contrast to less than $0.004$ percent in the satellite galaxies with $M_{*}\sim 10^9 M_{\odot}$ in the Auriga-L3.  

Note that, the mass fraction of PSR is unity in about $16$ percent satellite galaxies in the Auriga-L3 galaxies. These systems have early star formation with an average star formation redshift of $z\approx6.6$. After carefully examining the evolution of these systems, we find that they only experienced a single star formation event and subsequently ceased further star formation due to feedback and UV background heating. The stars in these satellite galaxies are thought to be enriched in alpha elements ($\rm [\alpha/Fe]\sim 0.35$) and show no significant scatter \citep[e.g.][]{2012ApJ...759..115F}. However, due to the lack of early metal production and mixing processes in the Auriga simulation, we are unable to predict the detailed elemental abundance patterns of stars in these satellite galaxies. 

It is noteworthy that some satellite galaxies, even those with masses greater than $10^8 M_{\odot}$, do not contain any PSR, as shown at the bottom of Fig.~\ref{fig:satellite}. About 16 percent of the satellite galaxies contain no PSR in the Auriga-L3 galaxies, suggesting that their host dark matter halos have been polluted by surrounding galaxies before their first in-situ star formation episodes. These satellite galaxies have later star formation with an average star formation redshift of $z\approx4.2$. 

Although the quantitative PSR fraction values in these satellite galaxies depend on numerical resolution and the adopted sub-grid physics models, the qualitative results remain largely model independent, as shown in Fig.~\ref{fig:satellite}. Therefore, our results suggest that the low mass and old satellite galaxies are the best targets for searching for the PSR.

\subsection{Comparison with previous studies}

In this section, we compare our results with previous studies that investigated the spatial distribution of second generation stars or truly Pop III stars in Milky Way-sized halos using various simulation approaches. These studies adopted different definitions for such stars, modeling techniques, and resolutions, which may lead to divergent conclusions. Nevertheless, a systematic comparison helps clarify the differences, and highlights how our higher resolution simulations and sample selection contribute to a more refined understanding of the distribution of the PSR.

\cite{2006ApJ...653..285S} employed the  particle tagging method in an N-body simulation with mass resolution similar to Auriga-L4, to study the spatial distribution of first and second generation stars. In their fiducial model, the fraction of second-generation stars at different radii is approximately $10^{-3.5}$, with slightly higher fractions in the outer region compared to the inner region. However, our results show that this fraction is even lower, and the fraction of stars located in the bulge is smaller than that in the stellar halo, rather than being comparable.

\citet{2010MNRAS.403.1283G} performed a simple semi-analytical modeling of Pop III stars and first galaxies within Milky Way-like halos using high-resolution N-body simulations \citep[][]{2008MNRAS.391.1685S}. They found that half of the PSR reside within about 40 kpc of the Galactic center and emphasized that the surviving first galaxies are promising locations for finding such relics. Compared to our results, both studies predict a broadly similar radial distribution of PSR.

{\cite{2007ApJ...661...10B} analyzed smoothed particle hydrodynamical simulations of Milky Way analogues with gas mass resolution $m_{\rm gas} \sim 10^6 M_{\odot}$, and found that metal free stars are dispersed throughout the entire simulated galaxy at $z=0$, suggesting a higher likelihood of detecting them in the outer galactic regions. Their qualitative results regarding the spatial distribution of such metal-free stars are consistent across the four models they employed and also agree with those obtained in this work. These results suggest that the qualitative features of their spatial distribution primarily reflect the hierarchical merger origin of these stars in the central galaxy. However, the total mass of metal free stars they predicted is substantially larger than that obtained in this work, likely due to the lower numerical resolution in their simulations.

\cite{2017MNRAS.465.2212S} analyzed the present-day spatial distribution of metal free star particles in the APOSTLE simulation and found that about $75\%$ are located beyond 8 kpc from the Galactic center. It is broadly consistent with our results, although we found a slightly more extended distribution (about $85\%$). However, they do not provide direct information on the masses and kinematics of such particles. In addition, they showed the differences between the oldest stars and the metal free stars. This is consistent with our finding that PSR have a wide range of formation times.

\cite{2023MNRAS.519..483C} investigated the spatial distribution of extremely metal-poor stars ([Fe/H]$<-3$) in Milky Way analogues at $z=0$ using the TNG50-1 simulation. While they also performed a kinematic orbital decomposition, they did not examine the formation times of such stars. Owing mainly to differences in sample definitions, their quantitative results differ from ours. Nevertheless, the qualitative trends are consistent: our sample is likewise preferentially located in the stellar halo and in low-mass satellite galaxies. Their simulation has a mass resolution comparable to Auriga-L4; in this study, we further assess resolution dependence by employing the higher-resolution Auriga-L3 runs.

\section{Conclusion}

In this work, we have used the cosmological magnetohydrodynamic galaxy formation simulation Auriga to explore the total mass, spatial distribution, and kinematics of the Pop III star relics (PSR) in Milky Way analogues. Our main findings are summarized as follows.

\begin{enumerate}

 \item The formation time of PSR in the Auriga simulation spans a wide redshift range, from about $z=22$ to $z=4$. The averaged formation redshift of PSR is about $z=10.0$, and the total mass fraction of PSR is tiny, about $2\times10^{-5}$ of the entire stellar population in Auriga-L3.  

 \item The PSR tend to concentrate in the central region of the Auriga halos but are substantially less concentrated than the overall galactic stellar population. About half of the PSR reside within the $40$ kpc of the galactic center.

 \item Based on the orbital decomposition in the Auriga-L3, only a small fraction (about $4$ percent) of PSR end up in the bulge component, about $8$ percent PSR reside in the disk, and about $70$ percent the PSR are distributed sparsely as halo stars, in particular the outer stellar halo, which contains more than $46$ percent of the PSR. 
 
 \item The present day satellite galaxies contain significant fraction of PSR, accounting for about $18$ percent of the total PSR in Auriga-L3. The PSR mass fraction of a satellite galaxy correlates strongly with its total stellar mass $M_{*}$, with lower mass satellite galaxies containing a higher fraction of PSR.

\end{enumerate}

Our results suggest that the outer stellar halo and low mass faint satellite galaxies are the most promising sites to search for PSR. Since most of these stars are located in the outer regions of the Milky Way, the next generation of deeper survey projects, such as Chinese Space station Survey Telescope (CSST), will help to greatly expand the PSR sample size. While some quantitative aspects of our results may be affected by numerical resolution, the qualitative trends are largely robust. Therefore, the results presented in this paper serve as a valuable guide for identifying the PSR in our Galaxy and, ultimately, contribute to constraining the properties of Pop III stars.

\section{Acknowledgments}
\begin{acknowledgments}

HY are grateful to Shihong Liao for their useful discussions and comments. We acknowledge the supports from the National Natural Science Foundation of China (Grant No. 12588202) and the National Key Research and Development Program of China (Grant No. 2023YFB3002500). QG acknowledges the hospitality of the International Centre of Supernovae (ICESUN), Yunnan Key Laboratory at Yunnan Observatories Chinese Academy of Sciences, and European Union’s HORIZON-MSCA-2021-SE-01 Research and Innovation programme under the Marie Sklodowska-Curie grant agreement number 101086388.

\end{acknowledgments}

\vspace{5mm}



\bibliography{sample631}{}
\bibliographystyle{aasjournal}

\end{document}